\documentstyle[12pt,epsf]{article}
\newcommand{\LU}{\lambda_U}

\newcommand{\LD}{\lambda_D}

\newcommand{\LL}{\lambda_L}

\newcommand{\gev}{\,{\rm GeV}}
\newcommand{\tev}{\,{\rm TeV}}

\newcommand{\lsim}{
  \raisebox{0.2em}{$<$} \hspace{-0.75em} \raisebox{-0.2em}{$\sim$} }

\newcommand{\KK}{ K^0 \! - \! \overline{K} {}^0 }

\newcommand{\BdBd}{ B^0_d \! - \! \overline{B} {}^0_d }
\title{ \vspace*{-1cm}
\begin{flushleft}
\hspace*{10cm} {\normalsize KEK-TH-523} \vspace{-3mm} \\
\hspace*{10cm} {\normalsize KEK Preprint 97-90}\vspace{-3mm} \\
\hspace*{10cm} {\normalsize June 1997} \\
\end{flushleft}
\mbox{} \vspace{10mm} \\
\bf Effect of SUSY phases on the $\BdBd$ mixing
 in the minimal supergravity model} 
\author{ }
\date{ }
\begin{document}
\baselineskip 6mm
\renewcommand{\thefootnote}{\fnsymbol{footnote}}
\maketitle
\vspace*{-1cm}
\begin{center}
{\large \sc Takeshi Nihei }\footnote{JSPS Research Fellow.} \\
 \mbox{} \vspace{3mm}\\
 {\em  Theory Group, KEK, Tsukuba, Ibaraki 305, Japan}
\end{center}
\thispagestyle{empty}
\vspace{0.7cm}
\begin{abstract}
\normalsize 
We investigate the effect of SUSY phases ( $\theta_A$ 
and $\theta_\mu$ ) on the $\BdBd$ mixing in the minimal supergravity
 model. It is known that the complex phase $\theta_A$ $=$ arg($A$) 
( $A$ is the universal coefficient of the trilinear scalar couplings ) 
is essentially unconstrained by the electric dipole moment experiment, 
while the phase $\theta_\mu$ $=$ arg($\mu$) 
( $\mu$ is the supersymmetric Higgsino mass )
 is strongly constrained to zero. 
We found that $\theta_A$ does not affect the phase of the $\BdBd$ mixing 
matrix element $M_{12}(B)$ by numerical analysis of the renormalization 
group equations. 
This means 
that the measurement of the $\BdBd$ mixing at the future 
B-factory could give the direct information on the parameters of
the CKM matrix even in the framework of the minimal supergravity model 
with the SUSY phase $\theta_A$. 
\end{abstract}
\newpage
\setcounter{page}{1}
%
%
{\it 1. Introduction} \\ 
\hspace*{0.5cm}
The minimal supergravity (SUGRA) model\cite{SUGRA} 
is expected to be the physics beyond the standard model. 
Due to the supersymmetry (SUSY), the quadratic divergence to the scalar
$({\rm mass})^2$ cancels out and it helps the theory with elementary 
scalar fields to be natural. 
Furthermore 
the spontaneous breaking of the SUGRA can provide the preferable structure
of soft SUSY breaking terms.

According to the current SUSY particle searches, masses of 
these particles are considered to be rather large. 
Even if the SUSY particles are too heavy to decay at presently working 
colliders, they may be detected through their
radiative effects. 
Hence indirect tests for SUSY models are important. 

CP violation in the neutral meson mixing 
is one of such indirect processes. Here we
focus on the CP violation in the $\BdBd$ mixing, which is one of the 
main targets of B-factory experiments. In this case the effects of new 
physics can be extracted from arg[$M_{12}(B)$], where $M_{12}(B)$ is 
the $\BdBd$ mixing matrix element. 
In general it seems that arg[$M_{12}(B)$] depends on SUSY parameters. 
If so, we cannot directly obtain the informations on the 
Cabibbo-Kobayashi-Maskawa (CKM) matrix. 

The prediction of the minimal SUGRA 
model for arg[$M_{12}(B)$] has been analysed\cite{arg M12B}
\cite{BB_LR}\cite{BB_LR_QCD}\cite{GNO}.
The result is 
that arg[$M_{12}(B)$] is the same as the standard model prediction 
independent of SUSY parameters. 
In these analyses, however, 
the SUSY parameters at the Planck (or GUT) scale are 
assumed to be real. This is because it seems natural for these parameters
to be real in order to suppress the electric dipole moments (EDMs) of the 
neutron and the electron. 
On the other hand, it is found that it is possible for the SUSY parameter
$A$ ( the universal coefficient of the trilinear scalar couplings ) to have 
a complex phase of order one\cite{EDM from two phases}. 
On the $\KK$ mixing it has already shown that the phase of $A$ 
does not change the phase of the matrix element $M_{12}(K)$ 
in the previous analysis\cite{Dugan et.al.} 
where a mass insertion approximation is adopted. 

In this letter we make an analysis of arg[$M_{12}(B)$] in the case of 
the complex $A$ parameter. We have solved the renormalization group 
equations (RGEs) numerically including all the off-diagonal elements of 
the Yukawa coupling matrices\cite{Bouquet et.al.}\cite{GNA}\cite{GNO}, 
while they have been ignored in most of the previous works. 
Some phenomenological constraints on the SUSY parameters are considered. 
We take the effects of the right-handed 
external bottom quarks into account in evaluating $M_{12}(B)$. 
QCD corrections below the weak scale are included. 
Implement for B-factory measurement is also mentioned. 

\vspace{1cm} 	
%
%
\hspace*{-0.6cm}{\it 2. The SUSY phases in the minimal SUGRA model} \\
\hspace*{0.5cm}  
We examine the low energy effective theory of the minimal SUGRA 
model with    
chiral superfields for three generations of quarks 
( $Q_i$, $U^c_i$ and $D^c_i$ ) and leptons 
( $L_i$ and $E^c_i$ ), chiral superfields for two Higgs doublets 
( $H_1$ and $H_2$ ), and vector superfields for 
the gauge group 
${\rm SU(3)_C}$ $\times$ ${\rm SU(2)_L}$ $\times$ ${\rm U(1)_Y}$. 
The superpotential is written as follows:
\begin{equation}
W = H_2 U^c \LU Q + H_1
D^c \LD Q
+ H_1 E^c \LL L + \mu H_1 H_2, 
\label{eqn:superpot of MSSM}
\end{equation}
where $\LU$, $\LD$ and $\LL$ are Yukawa coupling matrices in the 
generation space. 
The generation indices ($i$ $=$ 1, 2, 3) are suppressed.

The general soft SUSY breaking consists of the following terms: 

\begin{itemize}
\item[(i)]
scalar masses: \\
$ \tilde{q}_L^{\dagger} M_Q^2 \tilde{q}_L
 + \tilde{u}_R^{\dagger} M_U^2 \tilde{u}_R
 + \tilde{d}_R^{\dagger} M_D^2 \tilde{d}_R 
 + \tilde{l}^{\dagger}_L M_L^2 \tilde{l}_L
 + \tilde{e}_R^{\dagger} M_E^2 \tilde{e}_R
 + M_{H_1}^2 | h_1 |^2 + M_{H_2}^2 | h_2 |^2 $. 
\item[(ii)]
A-terms: 
$h_2 \tilde{u}_R^\dagger A_U
\tilde{q}_L + h_1 \tilde{d}_R^\dagger A_D \tilde{q}_L
 + h_1 \tilde{e}_R^\dagger A_L \tilde{l}_L + {\rm h.c.}$. 
\item[(iii)]
B-terms: $B \mu h_1 h_2 + {\rm h.c.}$. 
\item[(iv)]	
gaugino masses: 
$ \frac{1}{2} M_1 \tilde{\lambda}_1 \tilde{\lambda}_1 
+ \frac{1}{2} M_2 \tilde{\lambda}_2 \tilde{\lambda}_2 
+ \frac{1}{2} M_3 \tilde{\lambda}_3 \tilde{\lambda}_3 + {\rm h.c.} $. 
\end{itemize}
Here $\tilde{q}_L$, $\tilde{u}_R^\dagger$, $\tilde{d}_R^\dagger$, 
$\tilde{l}_L$, $\tilde{e}_R^\dagger$, $h_1$ and $h_2$ are the scalar
components of $Q$, $U^c$, $D^c$, $L$, $E^c$, $H_1$ and $H_2$, 
respectively. 
The fields $\tilde{\lambda}_\alpha$ ( $\alpha$ $=$ 1, 2, 3 ) 
denote gauginos. 

These soft SUSY breaking terms 
 which result from the couplings 
to the hidden sector of $N=1$ SUGRA have universal structure
at the Planck scale ($M_P$ $\sim$ $10^{19} \gev$), 
if we assume that the hidden sector is 
flavor-blind. 
In this analysis we put the following
boundary conditions at the GUT scale ($M_X$ $\sim$ $10^{16} \gev$) 
for simplicity, ignoring the 
RGE running effects between $M_P$ and $M_X$: 
\begin{itemize}
\item[(i)]
universal scalar masses: \\
 $M_Q^2=M_U^2=M_D^2=M_L^2=M_E^2=m_0^2 {\bf 1}$,
$M_{H_1}^2=M_{H_2}^2=m_0^2$,
\item[(ii)]
universal A-terms: $A_U=A m_0 \LU$, $A_D=A m_0 \LD$,  $A_L=A m_0 \LL$,
\item[(iii)]
universal gaugino masses: $M_1$ $=$ $M_2$ $=$ $M_3$ $=$ $M_g$.
\end{itemize}
These universal structures are required in order to suppress 
the flavor changing neutral current processes. 
The last relation for gaugino masses is derived if we assume the 
supersymmetric grand unification\cite{SUSY GUT}, 
which is strongly suggested from the precise 
measurements of the gauge coupling constants 
at LEP\cite{gauge coupling unification}. 
In our analysis we assume the realization of the grand unification
which is consistent with the negative results of the proton decay
experiments in the wide parameter region, though we don't specify
the unified gauge group\footnote{
Therefore we have not included the proton
decay analysis in this letter. }. 

It is known that the low energy effective theory of the 
minimal SUGRA model has four physical
 phases\cite{Dugan et.al.}: (i) the phase $\delta_{\rm CKM}$
 in the CKM matrix, 
(ii) the phase $\theta_A$ $=$ ${\rm arg}(A)$, 
(iii) the phase $\theta_\mu$ $=$ arg($\mu$) and 
(iv) the QCD vacuum parameter $\overline{\theta}_{\rm QCD}$. 
Here we have taken such a convention that $B \mu$ is real at
 the weak scale by phase rotation of the Higgs fields. 
Then the vacuum expectation values of the Higgs fields $h_1$ and $h_2$
 are found to be real. 
The universal gaugino mass $M_g$ is made real by an R-rotation. 
Throughout our analysis we assume $\overline{\theta}_{\rm QCD}$ $=$ 0.
The two phases $\theta_A$ and $\theta_\mu$ are peculiar to SUSY models, 
hence we call them 'SUSY phases'. 

Breaking of the ${\rm SU(2)_L}$ $\times$ ${\rm U(1)_Y}$ gauge symmetry 
is realized radiatively through the large Yukawa coupling constant of the 
top quark\cite{radiative breaking}. 
At the weak scale we minimize the Higgs potential to determine $|\mu|$ and 
$|B|$ by the condition that the vacuum expectation values 
of the neutral Higgs 
bosons ( $\langle h_1^0 \rangle$ $\equiv$ $v_1$ and $\langle h_2^0
\rangle$ $\equiv$ $v_2$ ) give the correct weak boson mass by 
$m_W^2$ $=$ $g_2^2(v_1^2+v_2^2)/2$. 

\vspace{1cm} 
%
%
%
\hspace*{-0.6cm}{\it 3. Constraint on the SUSY phases from 
EDM experiments} \\
\hspace*{0.5cm}
%
Nonvanishing particle EDMs are indications of CP violation. The current 
experiments for EDMs give stringent limits especially for the neutron 
and the electron: 
$|d_n| \, \lsim \, 1 \times10^{-25} e \cdot {\rm cm}$\cite{exp of NEDM}
 and 
$|d_e| \, \lsim \, 1 \times10^{-26} e \cdot {\rm cm}$\cite{exp of EEDM}, 
respectively. 
In principle these bounds put severe constraints on 
$\theta_A$ and $\theta_\mu$.  
However it was found that $\theta_A$ is essentially unconstrained, 
while $\theta_\mu$ is strongly constrained 
to be vanishing\cite{EDM from two phases}.
The reason is as follows. 
The EDMs receive three contributions: (i) chargino-squark loop, (ii) 
gluino-squark loop and (iii) neutralino-squark loop. The results of 
EDM calculation are given in 
Ref.\cite{EDM in SUGRA}\cite{Kizukuri & Oshimo}. 
It was found that the chargino contribution $d^{(C)}$ is dominant in the 
minimal SUGRA model\cite{Kizukuri & Oshimo}.  
The gluino contribution $d^{(G)}$ 
is subdominant and 
the neutralino contribution $d^{(N)}$ is the smallest. 
With the relation $d^{(C)}$ $\sim$ Im($\mu$), the phase $\theta_\mu$ 
is strongly bound to zero. On the other hand $\theta_A$ comes in the
subdominant gluino contribution $d^{(G)}$ $\sim$ Im($Am_0+\mu \tan \beta$), 
therefore $\theta_A$ does not have such a severe constraint. 
These statements on $\theta_A$ and $\theta_\mu$ hold independent 
of $\delta_{\rm CKM}$ because the diagrams exchanging the first 
generation of squarks/sleptons are dominant. 

\begin{figure}[t]
\begin{center}
\epsfxsize=15cm
\leavevmode\epsffile{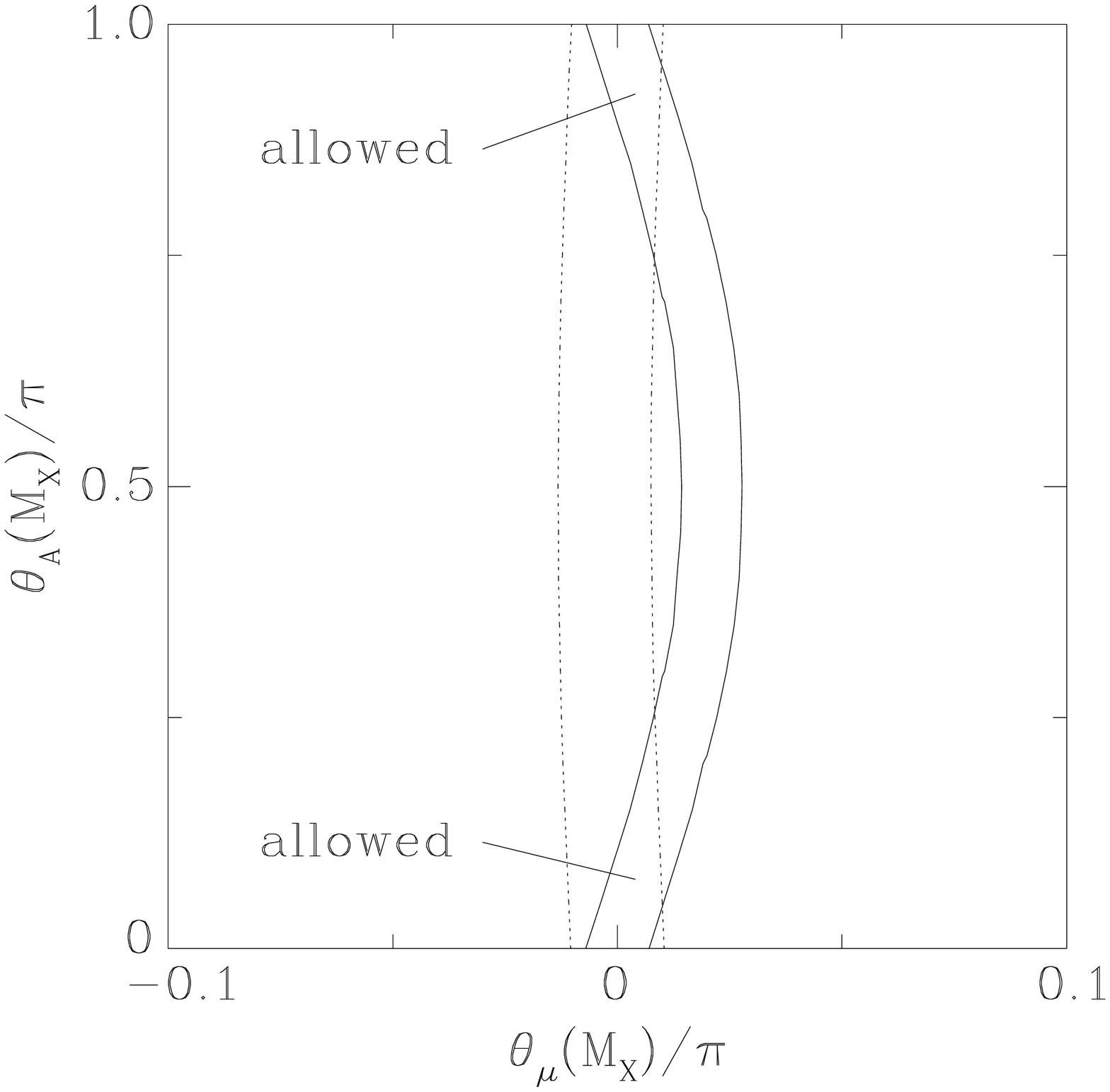}
\end{center}
\caption[aa1]{The allowed region on the ( $\theta_\mu$,  $\theta_A$ )
plane from the experimental constraints of $d_n$ and $d_e$
 in the minimal SUGRA model. 
The parameters except these phases are fixed as 
$|A| = 0.5$, $m_0 = 300 \gev$, $M_g = 100 \gev$ and $\tan \beta =3$. 
The region to satisfy the bound $|d_n|$ $<$ $1 \times10^{-25} e 
\cdot {\rm cm} $ is shown 
 between the two solid lines. The region to satisfy the bound 
$|d_e|$ $<$ $1 \times10^{-26} e \cdot {\rm cm} $
 is shown between the two dashed lines. 
The allowed regions are obtained by combining the two constraints. 
We have denoted 
$\theta_\mu$ as $\theta_\mu(M_X)$. 
 This is because the phase of $\mu$ does not run 
as seen from the RGE of the $\mu$ parameter: $\dot{\mu}$ $\sim$ 
(real)$\times$ $\,\mu$. 
On the other hand, the phases of $A_U$, $A_D$ and $A_L$ run due to
the contributions of the gaugino masses. }
\label{fig:dndeA}
\end{figure}

\begin{figure}[t]
\begin{center}
\epsfxsize=15cm
\leavevmode\epsffile{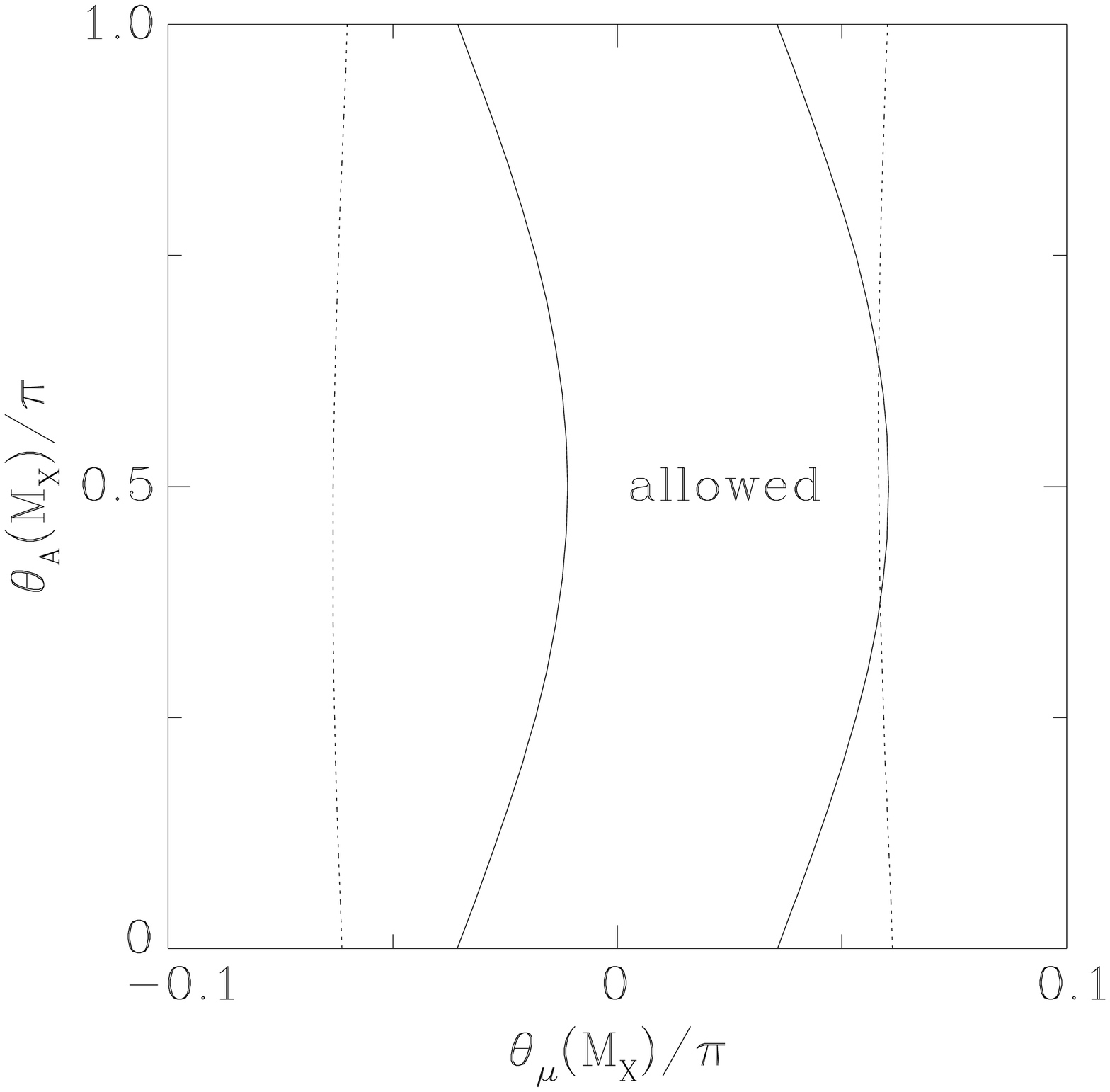}
\end{center}
\caption[aa1]{The same as Fig. \ref{fig:dndeA} for $|A| = 0.5$, 
$m_0 = 700 \gev$, $M_g = 300 \gev$ and 
 $\tan \beta =3$. }
\label{fig:dndeB}
\end{figure}

We have confirmed the statement in Ref.\cite{EDM from two phases}. 
In Fig. \ref{fig:dndeA} we present the allowed region 
on the ( $\theta_\mu$,  $\theta_A$ ) plane, where we have fixed 
the parameters as $|A| = 0.5$, $m_0 = $ $ 300 \gev$, $M_g =$ $100 \gev $
and $\tan \beta$ $\equiv$ $v_2/v_1$ $=3$.
We take the CKM parameters as
$|V_{us}|=0.221$, $|V_{cb}|=0.041$, $|V_{ub}|/|V_{cb}|=0.08$ and
$ \delta_{\rm CKM}$ $=$ $\pi/3$ in 
the standard parametrization\cite{PDG}. 
The result is almost independent of these CKM parameters. 
The region to satisfy the $d_n$ bound 
$|d_n| < 1 \times10^{-25} e \cdot {\rm cm} $ is shown 
 between the two solid lines. The region to satisfy the $d_e$ bound 
$ |d_e| < 1 \times10^{-26} e \cdot {\rm cm} $
 is shown between the two dashed lines. The allowed regions are obtained by
 combining the two constraints. 
This figure shows that $\theta_{\mu}$ has the strong constraint: 
$\theta_{\mu}$ $\lsim$ $0.01 \pi$, while $\theta_A$ 
does not have such a strong constraint around the small $\theta_{\mu}$ region. 
Figure \ref{fig:dndeB} is a similar result for $|A| = 0.5$,
$m_0 = $ $ 700 \gev$, $M_g =$ $300 \gev $ and $\tan \beta = 3$. 
In this case we don't have any constraint on $\theta_A$ around 
the small $\theta_{\mu}$ region. 

%

\vspace{1cm} 
%
%
\hspace*{-0.6cm}{\it 4. Effect of $\theta_A$ on the phase of $M_{12}(B)$} \\
\hspace*{0.5cm}
%
From the above discussion, it follows that the phase $\theta_A$ may be 
large in the small $\theta_\mu$ region. 
Now we wish to examine whether a large $\theta_A$ can affect 
the complex phase of $M_{12}(B)$. 

The matrix element $M_{12}(B)$ is estimated 
by the usual box diagram calculation. 
In the minimal SUGRA model there are five contributions to $M_{12}(B)$: 
\begin{itemize}
\item[(i)] W-boson and up-type quark loop (SM),
\item[(ii)] charged Higgs and up-type quark loop,
\item[(iii)] chargino and up-type squark loop,
\item[(iv)] gluino and down-type squark loop,
\item[(v)] neutralino and down-type squark loop or
neutralino, gluino and down-type squark loop. 
\end{itemize}
Among these the neutralino contribution (v) is expected to be much
smaller than the gluino contribution (iv) due to smallness of gauge coupling
constants and down type Yukawa coupling constants. Therefore we
neglect it in this analysis. 
We have included the diagrams with the 
external right-handed bottom quarks, though they are subdominant.
An analytic expression for $M_{12}(B)$ is found 
in Ref.\cite{BB_LR} and we 
have also considered QCD corrections following the method described
in Ref.\cite{BB_LR_QCD}. 

At first we give a brief summery in the case of $\theta_A$ $=$ 
$\theta_\mu$ $=0$. In this case it is known that arg[$M_{12}(B)$]
in the minimal SUGRA model is the same as the prediction of the 
standard model\cite{arg M12B}: arg[$M_{12}(B)$]$|_{\rm SUGRA}$ $=$ 
arg[$M_{12}(B)$]$|_{\rm SM}$ $=$ arg[$\xi_t^2$], where $\xi_i$ $\equiv$ 
$V_{id}^{*}V_{ib}$ ($i$ $=$ $u$, $c$ and $t$). 
The reason is that due to the degeneracy between the first two 
generations of squarks their contributions sum up to the terms 
proportional to $\xi_t(\xi_u+\xi_c)$ or $(\xi_u+\xi_c)^2$, which 
are equal to $-\xi_t^2$ and $\xi_t^2$ respectively because of the 
unitarity of the CKM matrix. 

Now we consider the effect of $\theta_A$ and present our
numerical result. In the following analysis,
 the phase $\theta_\mu$ 
is fixed at zero in order to satisfy the EDM constraints. 
We investigate the three dimensional parameter space $\{ |A|, m_0, M_g
\}$ in the range of $-5<|A|<5$, $0<m_0<2 \tev$ and $\ 0<M_g<2 \tev $. 
Moreover we require the phenomenological constraints in the following: 
\begin{itemize}
\item[(i)] The mass of any charged superparticle is larger than 45
  $\gev$\cite{LEP 1pre}\cite{LEP 1}. 
\item[(ii)] The gluino mass is larger than 100 $\gev$\cite{gluino mass 
  bound}. 
\item[(iii)] All the sneutrino masses are larger than 41
  $\gev$\cite{PDG old}.  
\item[(iv)] The CLEO result for the $ b \rightarrow s \gamma $
  inclusive branching ratio : 1 $\times$ $10^{-4}$ $<$ ${\rm
  Br}(b \rightarrow s \gamma)$ $<$ 4.2 $\times$
  $10^{-4}$\cite{CLEO}\cite{b_s_gamma}.
\item[(v)] The bounds from the neutralino search at LEP on the decay
  width $\Gamma(Z^0 \rightarrow \chi \chi)$ $<$ 8.4 MeV and the
  branching ratios ${\rm Br}(Z^0 \rightarrow \chi \chi^\prime)$,
 ${\rm Br}(Z^0 \rightarrow \chi^\prime \chi^\prime)$, $<$ 2 $\times$
 $10^{-5}$, where $\chi$ is the lightest neutralino and $\chi^\prime$
 is any neutralino other than $\chi$\cite{bound for
 neutralino}\cite{LEP 1}.
\item[(vi)] The lightest superparticle (LSP) is neutral\cite{LSP is
  neutral}. 
\item[(vii)] The condition to avoid the color and charge breaking
  vacua\cite{C&EM unbroken}. 
\end{itemize}

\begin{figure}[t]
\epsfxsize=15cm
\leavevmode\epsffile{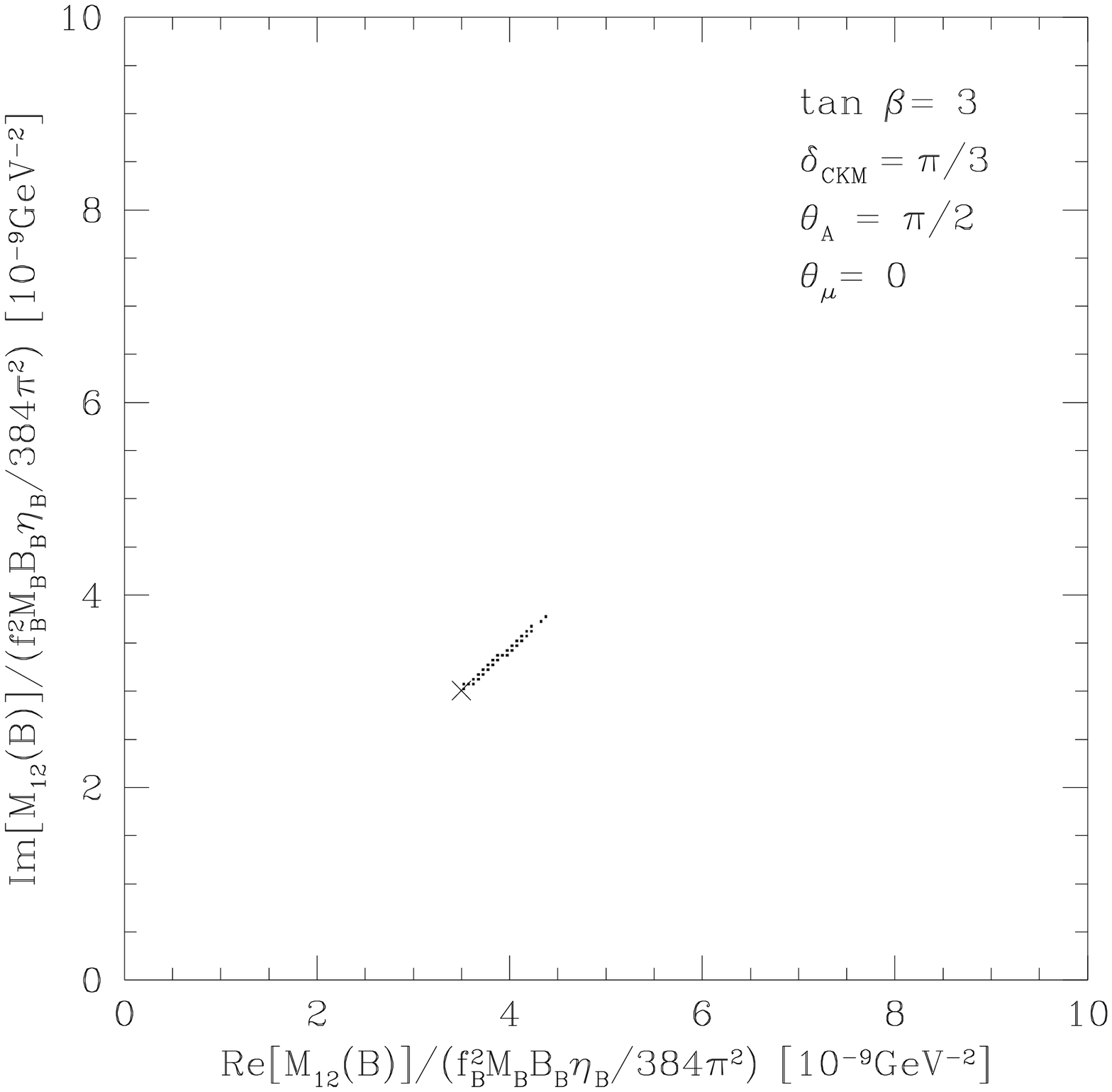}
\caption[aa111]{
The complex value of $M_{12}(B)$ in the minimal SUGRA
model. 
The SUSY phases are fixed as $\theta_A = 
 \pi/2$ and $\theta_{\mu}$ $= 0 $.
We have scanned the three dimensional parameter space $\{ \ |A|$, 
$m_0$, $M_g \ \}$ in the range of 
$-5<|A|<5$, $0<m_0<2 \tev$ and $0<M_g<2 \tev$. This is
the result for $\tan \beta = 3$ and $\delta_{\rm CKM}$ $=$ $\pi/3$. 
The cross represents the standard model prediction. 
The axes are normalized so as to remove the uncertainty 
of low energy hadron physics. The constants $f_B$, $M_B$ and $B_B$ are 
the decay constant, the mass and the bag parameter of the B-meson 
respectively. The constant $\eta_B$ is the QCD correction factor: 
$\eta_B$ $=$ $\alpha_3(m_W)^{6/23}$. }
\label{fig:m12com1}
\end{figure}

In Fig. \ref{fig:m12com1}, we show the complex quantity $M_{12}(B)$
normalized so as to remove the uncertainty of low energy hadron
physics. 
In this figure we take $\tan \beta = 3$ and we plot the SUGRA predictions
for allowed SUSY parameters. 
The cross represents the standard model prediction. 
The SUSY phases are fixed as $\theta_A$ $=$ $\pi/2$ and $\theta_\mu$ $=$ 0. 
We have fixed the CKM parameters as
$|V_{us}|=0.221$, $|V_{cb}|=0.041$, $|V_{ub}|/|V_{cb}|=0.08$ and
$ \delta_{\rm CKM}$ $=$ $\pi/3$. 
The top quark mass is fixed at $m_t=175 \gev$\cite{m_top}. 
One finds from Fig. \ref{fig:m12com1} that the complex phase of 
$M_{12}(B)$ is not affected by the phase $\theta_A$ as well as by 
the soft SUSY breaking parameters $|A|$, $m_0$ and $M_g$. 
We have also confirmed that 
the similar results are obtained for another choices of 
$\theta_A$ and $\tan \beta$. 
Though the phase of $M_{12}(B)$ depends, of course, 
on the CKM parameters, 
it holds even for another choice of the CKM parameters 
that the phase is the same as the corresponding 
standard model prediction. 

The reason for $\theta_A$ independence is as follows. 
The phase $\theta_A$ comes mainly 
through the squark left-right mixing 
in the box diagram calculation. 
Hence the diagrams exchanging the W-boson or the charged Higgs are
trivially independent of $\theta_A$. 
In the chargino contribution, it follows that
 the squark left-right mixing appears only in the combination of 
 $\sim$ $|Am_0+\mu \cot \beta|^2$. 
Therefore the phase cancels out and $\theta_A$ does not change 
the phase of $M_{12}(B)$. However $\theta_A$ can affect the absolute
value of $M_{12}(B)$ because $Am_0$ is only a part of 
the left-right mixing. 
As for the gluino contribution, 
the effect of $\theta_A$ is negligible
because the left-right mixing of the down-type squarks is quite small. 

We have also investigated the effect of $\theta_\mu$. 
From our numerical evaluations it is found that 
$\theta_\mu$ also does not affect arg[$M_{12}(B)$] even if we remove the 
EDM constraints and give a large complex phase to $\mu$. 

From our analysis we conclude that the SUSY phase $\theta_A$ does not
change the phase of the $\BdBd$ matrix element $M_{12}(B)$. 
Combined with the previous results for $\theta_A$ $=$ 
$\theta_\mu$ $=$ 0\cite{arg M12B}\cite{GNO}, it follows that 
arg[$M_{12}(B)$] is completely determined by the CKM parameters. 
It means that the measurement of CP asymmetry at the future
B-factory could give the direct information on the parameters of the
CKM matrix even in the framework of the minimal SUGRA model with the
SUSY phase $\theta_A$. 

%

%

\vspace{1cm} 
%
%
\hspace*{-0.6cm}{\it 5. Summery} \\
\hspace*{0.5cm}
In this letter we have examined the 
effect of the CP violating SUSY phase $\theta_A$ on the phase of
the $\BdBd$ mixing matrix element $M_{12}(B)$ in the 
minimal SUGRA model. 
We have solved the RGEs
numerically including all the off-diagonal elements of the 
Yukawa coupling matrices\cite{Bouquet et.al.}\cite{GNA}\cite{GNO}, 
while they have been ignored in most of the previous works. 
It is found that 
the phase $\theta_A$ does not 
change the phase of $M_{12}(B)$ and it is completely determined 
by the CKM parameters. 
This means 
that the measurement of the $\BdBd$ mixing at the future 
B-factory could give the direct information on the parameters of
the CKM matrix even in the framework of the minimal SUGRA model with
the SUSY phase $\theta_A$.  


\vspace{1cm}
%
%
\hspace*{-0.6cm}{\it Acknowledgement} \\
\hspace*{0.5cm} I would like to thank Y. Okada and T. Goto for useful
discussions. 
This work is supported in part by the Grant-in-Aid for Scientific
Research from the Ministry of Education, Science and Culture, Japan. 
%

%
%
%
%
\end{document}